\documentclass[12pt]{article}
\usepackage{jhep-mod}
\usepackage{bm}
\usepackage{amssymb, amsmath, amsthm, physics}
\usepackage{mathrsfs}
\usepackage{hyperref}
\usepackage{multicol}
\usepackage{microtype}
\usepackage{float, array, tabu}
\usepackage[tableposition=top]{caption}
\usepackage{tocloft}
\usepackage[utf8]{inputenc}
\usepackage{enumerate}
\usepackage{bigints}
\usepackage{appendix}
\usepackage{graphicx}
\usepackage{float}
\usepackage{tikz}
\usetikzlibrary{patterns}
\usepackage{setspace}
\usepackage{cancel}
\usepackage[normalem]{ulem}
\usepackage{doi}
\definecolor{purple}{rgb}{1,0,1}
\definecolor{lime}{HTML}{A6CE39} 

\definecolor{lime}{HTML}{A6CE39}
\newcommand{\orcidicon}{%
	\begin{tikzpicture}
	\draw[lime, fill=lime] (0,0) 
		circle [radius=0.16] 
		node[white] {{\fontfamily{qag}\selectfont \tiny ID}};
	\draw[white, fill=white] (-0.0625,0.095) 
		circle [radius=0.007];
	\end{tikzpicture}
	\hspace{-5mm}
}
\newcommand\orcidJessica{{\href{https://orcid.org/0000-0002-2669-2899}{\orcidicon}}}
\newcommand\orcidSebastian{{\href{https://orcid.org/0000-0003-1997-0026}{\orcidicon}}}
\newcommand\orcidMatt{{\href{https://orcid.org/0000-0003-1088-6485}{\orcidicon}}}
\begin{document}
\title{
\null\vspace{-75pt}
\huge Generic warp drives violate the \\
null energy condition
}
\author{
\Large
\leftline{Jessica Santiago$^{1,2}$\orcidJessica\!,
Sebastian Schuster$^3$\orcidSebastian\!, {\sf{and}}
Matt Visser$^1$\orcidMatt}}
\affiliation{
\vspace{-15pt}
$^1$ School of Mathematics and Statistics, 
Victoria University of Wellington, \\
\null\qquad 
PO Box 600, Wellington 6140, New Zealand.}
\affiliation{$^2$ Section of Astrophysics, Astronomy and Mechanics, 
Department of Physics,\\
\null\qquad 
Aristotle University of Thessaloniki, Thessaloniki 54124, Greece.}
\affiliation{$^3$ Institute of Theoretical Physics,
	Faculty of Mathematics and Physics, Charles University,\\
	\null\qquad V~Hole\v{s}ovi\v{c}k\'{a}ch~2,
	180 00 Prague 8,
	Czech Republic.}

\emailAdd{jessica.santiago@sms.vuw.ac.nz}
\emailAdd{sebastian.schuster@utf.mff.cuni.cz}
\emailAdd{matt.visser@sms.vuw.ac.nz}
\parindent0pt
\parskip7pt

\abstract{ Three recent articles have claimed that it is possible to, at least in theory,  either set up positive energy warp drives satisfying the weak energy condition (WEC), or at the very least, to minimize the WEC violations.  These claims are at best incomplete, since the arguments as presented only assert but do not prove  the existence of \emph{one} set of timelike observers, the co-moving Eulerian observers, who see relatively ``nice'' physics. While these particular observers might arguably see a positive energy density, the WEC requires \emph{all} timelike observers to see positive energy density. Therefore, one should carefully revisit this issue. A more careful analysis shows that the situation is actually much grimmer than advertised --- within the framework adopted by those three papers all physically reasonable warp drives will certainly violate the WEC, and both the strong and dominant energy conditions.  Under plausible subsidiary conditions the null energy condition is also violated.
While warp drives are certainly interesting examples of speculative physics, the violation of the energy conditions, at least within the framework of standard general relativity, is unavoidable. Even in modified gravity, physically reasonable warp drives will still violate the purely geometrical null convergence condition and the timelike convergence condition which, in turn, will place very strong constraints on any modified-gravity warp drive. 

\noindent

\smallskip
\noindent
{\sc Date:} Saturday 26 February 2022; 
\LaTeX-ed \today\\
(Closely resembles final version accepted and in press at PRD.)

Alcubierre warp drive; 
Nat\'ario generic warp drive; Nat\'ario zero-expansion warp drive; zero-vorticity warp drive.

}

\maketitle

\def\H{{\scriptscriptstyle{\mathrm{H}}}}
\def\ISCO{{\scriptscriptstyle{\mathrm{ISCO}}}}
\def\d{{\mathrm{d}}}
\def\O{{\mathcal{O}}}
\def\sign{{\mathrm{sign}}}
\def\L{{\mathcal{L}}}
\def\NEC{{\hbox{NEC}}}
\def\WEC{{\hbox{WEC}}}
\def\SEC{{\hbox{SEC}}}
\def\DEC{{\hbox{DEC}}}
\def\NCC{{\hbox{NCC}}}
\def\TCC{{\hbox{TCC}}}
\clearpage

\section{Introduction}\label{S:Intro}
Three recent articles~\cite{Lentz:2020, Bobrick:2021, Fell:2021} have argued for the existence of ``physically reasonable'' positive-energy warp drive configurations, either satisfying the weak energy condition (WEC), or minimizing violations thereof. (See also the series of articles~\cite{Osvaldo-1,Osvaldo-2,Osvaldo-3,Osvaldo-4,Osvaldo-5}.) These claims are in sharp contrast to the 25-year-old consensus opinion that at least some energy condition violations are necessary for the generation of warp fields~\cite{Alcubierre:1994, Lobo:2007, Everett:1997, Lobo:2004a, Everett:1995, VanDenBroeck:1999a, Natario:2001, Visser:1999, Hiscock:1997, Pfenning:1998, Lobo:2017,  Finazzi:2009, Lobo:2002, Clark:1999, Lobo:2004b, Low:1998, VanDenBroeck:1999b, McMonigal:2012, Alcubierre:2017, Alcubierre:2001,Olum:1998}. 

Specifically, known results already include:
\vspace{-15pt}
\begin{itemize}
\itemsep-3pt
\item Alcubierre warp drives have long been known to violate the WEC~\cite{Alcubierre:1994}, and have more recently been shown to also violate the null energy condition (NEC)~\cite{Alcubierre:2017}.
\item Nat\'ario zero-expansion warp drives have long been known to violate the WEC~\cite{Natario:2001}.
\item Generic Nat\'ario warp drives are known to violate the dominant energy condition (DEC)~\cite{Low:1998}, and have also been shown to violate either the strong energy condition (SEC) or the WEC~\cite{Natario:2001}.
\end{itemize}
\vspace{-10pt}
We shall strengthen these well known results below, first showing that the Nat\'ario zero-expansion warp drives also violate the SEC, and the NEC, and then ultimately showing that generic  Nat\'ario  warp drives (including the variants discussed in references~\cite{Lentz:2020, Bobrick:2021, Fell:2021}) violate the WEC, and under plausible subsidiary conditions also violate the NEC.  

How, then, is this result compatible with the various rather bold claims made in references~\cite{Lentz:2020, Bobrick:2021, Fell:2021}?  (And, for that matter, various parts of the popular press.) The key observation is that  WEC requires \emph{all} timelike observers to see positive energy density, whereas the analyses of references~\cite{Lentz:2020, Bobrick:2021, Fell:2021} at best only investigate the energy density as seen by \emph{one class} of timelike observers (the co-moving Eulerian observers). 
Thus the claims made in references~\cite{Lentz:2020, Bobrick:2021, Fell:2021} are at best grossly incomplete, and in many key specific details, wrong.

\enlargethispage{40pt}
To set the stage, in section~\ref{S:WarpDrives} we shall first discuss the kinematics of generic Natário warp drive spacetimes, and then calculate the relevant spacetime curvature (Riemann, Ricci, and Einstein tensors) in section~\ref{S:Curvature}. Assuming standard general relativity, we calculate in section~\ref{stress} all the components of the generic Natário warp-field stress-energy tensor. In section~\ref{S:ECIntro}, we then provide a number of general results regarding the implications of the point-wise energy conditions, and relate them to the null and timelike convergence conditions in section~\ref{S:CC}. With those preliminaries out of the way, the actual proof of energy condition violations in generic Natário warp-field spacetimes will be straightforward (section~\ref{S:ECs}). We conclude with some comments on the possibility of either moving beyond Einstein gravity, or the possibility of further generalizing the notion of warp-field spacetimes. Nevertheless, within the context of standard Einstein gravity, we emphasize that in physically reasonable warp field spacetimes energy condition violations are utterly unavoidable, and must be squarely faced. 

\clearpage
In the appendices, we discuss various technical issues arising in recent papers concerning themselves with warp drives: Appendix~\ref{S:spherical} deals with a confusion over the precise meaning of ``spherical symmetry'' in warp space-times. Appendix~\ref{A:B} points to misunderstandings regarding coordinate transformations. Appendix~\ref{S:defective} addresses issues of differentiability and related technical issues. Finally, Appendix~\ref{A:final} lists further technical oversights, small and large, of recent publications.

\section{Warp drive kinematics}\label{S:WarpDrives}

The generic Nat\'ario  warp drive line element we shall consider is this~\cite{Natario:2001}:
\begin{equation} \label{E:line}
     \d s^2 = - \d t^2 + \delta_{ij} \; \left( \d x^i - v^i(x,y,z,t)\, \d t\right) \; \left( \d x^j - v^j(x,y,z,t) \,\d t\right).  
\end{equation}
Nat\'ario calls this a ``warp drive spacetime''. Both the vector $v^i(x,y,z,t)$ \emph{and its derivatives} are assumed to be smooth and bounded~\cite{Natario:2001}. 
This line element is sufficiently general to cover well over 99\% of the relevant literature, and the very few exceptions in the extant literature will be explicitly discussed later on in appendix \ref{A:B}. Certainly the models discussed in references~\cite{Lentz:2020, Bobrick:2021, Fell:2021} fall into this generic Nat\'ario framework. 

The line element represents an ADM-like (3+1) decomposition of the metric~\cite{ADM1,ADM2,MTW,Wald,Poisson,Kiefer,Alcubierre:3+1}, with unit lapse $N\to1$; a flow vector $v^i = -$(shift vector);
and a flat spatial 3-metric $g_{ij} = \delta_{ij}$.
The sign-flip on the  shift vector is traditional in a warp drive context, and inspired by the notion of ``flow'' rather than ``shift''. Accordingly we shall speak of the ``flow vector'' rather than the ``shift vector''. (See also the traditional usage in the ``analogue spacetime'' programme~\cite{Unruh:1980,LRR,Visser:acoustic,
Novello:2002,Schutzhold:2002}.)
The explicit metric components are
\begin{equation}
g_{ab} = \left[ \begin{array}{c|c} -(1-v^2) & -v_j \\  \hline -v_i & \delta_{ij} \end{array} \right],
\qquad\qquad
g^{ab} = \left[ \begin{array}{c|c} -1 & -v^j \\ \hline -v^i & \delta^{ij} - v^i v^j\end{array} \right].
\end{equation}
The metric signature is manifestly $-+++$, while spacetime indices such as $a$, $b$, ... run from 0 to 3, and spatial indices such as $i$, $j$... run from 1 to 3. 
At large spatial distances we want the spacetime to be asymptotically flat, (that is, asymptotically Minkowski or at worst asymptotically Schwarzschild).

The most natural boundary conditions to impose are that $v^i(x,y,z,t)  \to 0$ at spatial infinity.  
One could alternatively impose $v^i(x,y,z,t)  \to v^i_*(t)$ at spatial infinity, with $v^i_*(t)$ some function of time only. This would still be asymptotically Minkowski, but with new spatial coordinates $x^i \to \bar x^i = x^i - \int v^i_*(t)\; \d t$. This would be useful, for instance, if one wishes to adopt a coordinate system moving with the warp bubble.  But there is certainly no loss of generality in enforcing $v^i(x,y,z,t)  \to 0$ at spatial infinity. 

\enlargethispage{30pt}
\clearpage
Indeed there is a very important issue of physics hiding here --- physically we want the warp bubble (for at least part of the time) to be ``moving'' with respect to the ``fixed stars'' --- otherwise the construction is physically uninteresting. Motion between the warp bubble and the fixed stars can be achieved thusly: Either by choosing coordinates such that $v^i(x,y,z,t)  \to 0$ at spatial infinity (so the fixed stars are ``at rest'') while the warp bubble has explicit time dependence, or alternatively,
by choosing coordinates where the warp bubble (the central region of the spacetime) is time independent and the fixed stars are in motion $v^i(x,y,z,t)  \to v^i_*(t)$ at spatial infinity. 

Furthermore, we want the warp bubble to be physically well-localized; we do not wish to have to accelerate the entire universe to get an interesting warp field. On the other hand, demanding that the flow vector have compact support (while mathematically convenient) is physically unnecessarily restrictive. For our purposes it will be more than sufficient for us to demand that the flow field falls off sufficiently rapidly near spatial infinity so that we can integrate by parts. The key point in proving NEC violations (and so implicitly proving violations of WEC, SEC, and DEC) is  the significantly weaker condition that gradients of the flow field tend to zero at spatial infinity.
We shall expand on this point more extensively in the discussion below. 

Following Nat\'ario, we shall also demand that the flow field (and hence the metric) be sufficiently smooth and bounded. At the very worst we shall allow Israel--Lanczos--Sen ``thin-shell'' distributional contributions to the Riemann tensor. Such assumptions are completely standard and implicit in the extant literature. 

Specifically, we shall demand  that the flow $\vec v$ (and hence the metric) be $C^{2-}$ at worst, that is,  piecewise  twice differentiable, with at worst discontinuous first derivatives. Then the Christoffel symbols are $C^{1-}$ at worst, 
that is,  piecewise  once differentiable with at worst step-function discontinuities. 
Then the Riemann tensor is $C^0$ at worst, 
 piecewise  continuous with at worst delta-function contributions. 

Finally, we emphasize the need for at least some subsidiary conditions to be applied to the generic Nat\'ario metric (\ref{E:line}) in order to distinguish a warp drive from, for instance,  the Painlev\'e--Gullstrand version of Schwarzschild spacetime. 
The line element (\ref{E:line})  is sufficiently general and explicit  to be physically interesting as a warp drive exemplar, but some subsidiary conditions would still be desirable.

\clearpage
Because the spatial 3-slices are intrinsically flat, and the lapse is unity, the only nontrivial physics hides in the extrinsic curvature of the spatial 3-slices, for which we have the particularly simple result
\begin{equation}
	\label{E: K}
K_{ij} = v_{(i,j)}.
\end{equation}
In order to prevent the warp field being trivial (Minkowski space) we will demand that there is at least one point in spacetime where the extrinsic curvature is nonzero. 

\enlargethispage{30pt}
Let us define $n_a = -\nabla_a t = (-1;0,0,0)_a$. 
Then we have $n^a = g^{ab} n_b = -g^{at} = (1,v^i)$. 
Thus $n^a$ is future-pointing. It is actually a future-pointing 4-velocity normal to the spatial 3-slices: 
$g^{ab} \,n_a n_b = -1 = g_{ab} \, n^a n^b$. 
Observers with 4-velocity $n^a$ are often called ``Eulerian''. 
In some sense (see discussion below) they ``go with the flow'', they are ``co-moving''. 
Furthermore, the Eulerian observers $n^a = - g^{ab} \, \nabla_b t$ are in fact timelike \emph{geodesics}. 
This is a simple consequence of the fact that the warp spacetimes are unit-lapse:
\begin{equation}
n^b \nabla_b n^a  = g^{ac} n^b \nabla_b \nabla_c t =  g^{ac} n^b \nabla_c \nabla_b t 
=  g^{ac} n^b \nabla_c n_b   
= {1\over2} g^{ac} \nabla_c( n^b n_b) = {1\over2} g^{ac} \nabla_c(-1) = 0.
\end{equation}

The warp drive spacetime is by construction globally hyperbolic, and at each and every instant in time ``$t$'' the  Eulerian observers  are 4-orthogonal to the flat spatial slices --- so the Eulerian observers define a zero-vorticity congruence of timelike geodesics that by construction \emph{cannot} have any focussing points.\footnote{Everett~\cite{Everett:1995} takes the point of view that one might take this warp drive metric as a local notion only, thus avoiding the global hyperbolicity conditions of the original Alcubierre space-time, and uses that to construct closed timelike curves (CTCs). This introduces a whole new level of complexity related to global causality and ``chronology protection'', of which more anon.
}

From the metric line element one can easily read off a suitable choice of co-tetrad: The timelike covector is simply 
\begin{equation}
(e^{\hat 0})_a = (1;0,0,0)_a =n_a, 
\end{equation}
while the spatial co-triad is
\begin{equation}
(e^{\hat 1})_a = (-v_x;1,0,0)_a, 
\quad (e^{\hat 2})_a = (-v_y;0,1,0)_a, 
\quad  (e^{\hat3})_a = (-v_z;0,0,1)_a. 
\end{equation}

\clearpage
The corresponding tetrad is then easily determined: The timelike leg is
\begin{equation}
(e_{\hat 0})^a = (1;v^x,v^y,v^z)^a= n^a, 
\end{equation}
while now the spatial triad is particularly simple
\begin{equation}
(e_{\hat1})^a = (0;1,0,0)^a, \quad (e_{\hat2})^a = (0;0,1,0)^a, \quad 
(e_{\hat3})^a = (0;0,0,1)^a. 
\end{equation}

This implies that, in the co-moving orthonormal basis, for any $T^0_2$ tensor:
\begin{equation}
X_{\hat a \hat b} = e_{\hat a}{}^a \; e_{\hat b}{}^b \; X_{ab} = 
\left[ \begin{array}{c|c} 
X_{\hat 0\hat0} & X_{\hat 0 \hat j} \\ \hline X_{\hat i \hat 0} & X_{\hat i \hat j}
\end{array} \right]
=
\left[ \begin{array}{c|c} 
X_{nn} & X_{nj} \\ \hline X_{in} & X_{ij}
\end{array} \right]
=
\left[ \begin{array}{c|c} 
X_{ab} n^a n^b & X_{aj} n^a \\ \hline X_{ia} n^a & X_{ij}
\end{array} \right].
\end{equation}
This is why objects such as  $X_{nn} = X_{ab} \, n^a \, n^b = X_{\hat0\hat0}$ and $X_{ni} = X_{ai} \, n^a= X_{\hat0\hat i}$ are so important. Subject to this choice of coordinates and tetrad, for the covariant spatial components we have the particularly simple result $X_{\hat i \hat j} = X_{ij}$. We will often use this observation to simplify formulae by suppressing the ``hats'' when they are not critical to understanding. 

\enlargethispage{20pt}
Similarly, for any fully covariant $T^0_4$ tensor one has 
$X_{\hat a\hat b\hat c\hat d} = e_{\hat a}{}^a \,e_{\hat b}{}^b \, e_{\hat c}{}^c \, e_{\hat d}{}^d  \, X_{abdc}$.  Then for any tensor that has the same symmetries as the Riemann tensor it suffices to calculate
\begin{equation}
X_{\hat0\hat i\hat 0 \hat j} = X_{ninj} = n^a n^b X_{aibj}; \qquad 
X_{\hat0\hat i\hat j \hat k} =  X_{nijk} = n^a X_{aijk}; \qquad
X_{\hat i\hat j \hat k\hat l} = X_{ijkl}. 
\end{equation}

Some specific examples of the generic warp drive spacetime are:
\vspace{-10pt}
\begin{description}
\itemsep-3pt
\item[Alcubierre warp field~\cite{Alcubierre:1994}:] \ \\
The original Alcubierre warp field~\cite{Alcubierre:1994}, taken to be moving in the $z$ direction with constant velocity $v_*$, is given by
 \begin{equation}
 \label{E:A1}
 v^i(x,y,z,t) = \left( 0,0, 1 \right)^i \; v_*\;  f\left(\sqrt{x^2+y^2+(z-v_* t)^2} \right).
 \end{equation}
 Here, $ f(0)=1$, and $ f(\infty)=0$.
 
 \enlargethispage{25pt}
 This was rapidly generalized~\cite{Alcubierre:1994} to a time-dependent velocity for the warp bubble
 \begin{equation}
 \label{E:A2}
 v^i(x,y,z,t) = \left( 0,0, 1 \right)^i \; v_*(t) \; f\left(\sqrt{x^2+y^2+\left(z- \int v_*(t)\,\d t \right)^2} \right).
 \end{equation}
 Note these specific models are ``spherically symmetric''. More on this point below.\\
Despite repeated assertions in reference~\cite{Bobrick:2021} there is absolutely no difficulty in making the velocity of the warp  bubble time-dependent. 
\\

\clearpage
\item[Nat\'ario zero-expansion warp field~\cite{Natario:2001}:] \ \\   
The Nat\'ario zero-expansion warp drive simply sets   $\div \vec v=0$.
Note that the second half of Nat\'ario's 2001 paper~\cite{Natario:2001} focusses on this zero-expansion warp field, whereas the first half of that paper deals with the generic Nat\'ario warp field spacetime of equation~(\ref{E:line}). 
When mentioning Natário's warp field, it is therefore important to make explicit which of these two specific warp drives one is referring to.
\\

\item[Lentz/Fell--Heisenberg zero-vorticity warp field~\cite{Lentz:2020,Fell:2021}: ] \ \\
  The Lentz/Fell--Heisenberg zero-vorticity warp drive uses a purely gradient flow $\vec v = \grad \Phi$, implying $\vec\omega = \curl \vec v=0$. 
\end{description}
\vspace{-5pt}
Thus the  generic warp drive line element (\ref{E:line}), introduced by Nat\'ario in~\cite{Natario:2001}, covers all three of these specific cases, and many more, and is sufficiently general to cover almost all of the relevant literature.

\section{Warp-field curvature }\label{S:Curvature}
The spacetime curvature for warp drive spacetimes is relatively easily determined via a specific application of the ADM formalism~\cite{ADM1,ADM2,MTW,Wald,Poisson,Kiefer,Alcubierre:3+1}.
\enlargethispage{20pt}
\subsection{Riemann tensor }
The Gauss--Codazzi equations would in general imply\footnote{See for instance equation (2.92) of Gourgoulhon~\cite{ADM2}.}:
\begin{equation} 
R_{\hat i\hat j\hat k\hat l} = {}^{(3)\!}R_{\hat i\hat j\hat k\hat l}
+   {K_{\hat i\hat k} K_{\hat j\hat l} - K_{\hat i\hat l} K_{\hat j\hat k}}.
\end{equation}
However, since our 3-geometry is flat, we have  $ {}^{(3)\!}R_{\hat i\hat j\hat k\hat l} \to 0$, and because of our choice of spatial triad we can dispense with the ``hats''. So for warp drive spacetimes we simply have
\begin{equation}
R_{ijkl} = K_{ik} K_{jl} - K_{il} K_{jk}.
\end{equation}
The Gauss--Mainardi equations imply\footnote{See for instance equation (2.101) of Gourgoulhon~\cite{ADM2}, specializing to $K_{ij} = v_{(i,j)}$.}:
\begin{equation}
R_{nijk} = R_{aijk} n^a =  K_{ij,k}-K_{ik,j} = v_{(i,j),k} -  v_{(i,k),j} =  v_{[j,k],i}.
\end{equation}

\clearpage
The remaining components of the Riemann tensor do not (to our knowledge) seem to have a special name, and are a little trickier to calculate. In the present context, to explicitly calculate $R_{ninj} $ one could either use brute force (\emph{e.g.,} {\sf Maple}), or appeal to a simplification of the ADM formalism,\footnote{For instance, use equation (3.43) of Gourgoulhon~\cite{ADM2}, specializing to $N\to 1$,\\ \null\qquad\quad and flipping the sign of the Lie derivative term to account for  (flow) $= -$(shift). } or use the general commutator identity 
\begin{equation}
[\nabla_a,\nabla_b] n_c = -R^d{}_{cab}\; n_d, 
\end{equation}
suitably projected onto spatial and normal $n$ directions.

One finds
\begin{equation}
R_{ninj} = R_{aibj} \,n^a \,n^b =  - \L_n K_{ij} + (K^2)_{ij}.
\end{equation}
Here $\L_n K_{ij}$ is the Lie derivative of the extrinsic curvature, where $\dot K_{ij}  = \partial_t K_{ij}$, and
\begin{equation}
\L_n K_{ij} = \dot K_{ij} + v^k \partial_k K_{ij} + \partial_i v^k K_{kj} +\partial_j v^k K_{ik}.
\end{equation}
Also
\begin{equation}
 (K^2)_{ij} = K_{ik} \,\delta^{kl}\, K_{lj}.
\end{equation}
We can rewrite the Lie derivative as 
\begin{equation}
\L_n K_{ij} = \dot K_{ij} + v^k \partial_k K_{ij} + \partial_{[i} v_{k]} K_{kj} 
+ K_{ik} \partial_{[j} v_{k]}+
2(K^2)_{ij},
\end{equation}
implying
\begin{equation}
\L_n K_{ij} = n^a \partial_a  K_{ij}  + \partial_{[i} v_{k]} K_{kj} 
+ K_{ik} \partial_{[j} v_{k]}+
2(K^2)_{ij}.
\end{equation}
Therefore
\begin{equation}
R_{ninj} = -n^a \partial_a  K_{ij}  - \partial_{[i} v_{k]} K_{kj} 
- K_{ik} \partial_{[j} v_{k]} - (K^2)_{ij}.
\end{equation}

\subsection{Ricci tensor }
Performing suitable contractions, for the Ricci tensor we find
\begin{equation}
R_{nn} = -\L_n K - \tr(K^2),
\end{equation}
where $\L_n K $ is the Lie derivative of $K= K_{ij} \, \delta^{ij} = \tr(K)$, the trace of the extrinsic curvature. That is,  $\L_n K = \dot K + v^i \partial_i K = n^a \,\partial_a K$, so we also have
\begin{equation}
R_{nn} = -n^a \partial_a  K - \tr(K^2).
\end{equation}
Furthermore, 
\begin{equation}
R_{ni} =   K_{ij,k} \; \delta^{jk}-K_{,i}.
\end{equation}
In view of the fact that in this context spatial indices are always raised and lowered using Kronecker deltas, we can and shall often simplify notation by contracting over two indices down and writing
\begin{equation}
R_{ni} =   K_{ij,j}-K_{,i}.
\end{equation}
Finally, for the spatial components
\begin{equation}
R_{ij} =   \L_n K_{ij}  + K K_{ij} - 2(K^2)_{ij},
\end{equation}
which can be recast as
\begin{equation}
R_{ij} =   n^a\partial_a K_{ij}  + K K_{ij} 
+ \partial_{[i} v_{k]} K_{kj} + K_{ik} \partial_{[j} v_{k]}.
\end{equation}
Alternatively
\begin{equation}
R_{ij} =   \partial_a ( n^a K_{ij}) 
+ \partial_{[i} v_{k]} K_{kj} + K_{ik} \partial_{[j} v_{k]}.
\end{equation}
Then for the Ricci scalar $R=-R_{nn} + \delta^{ij} R_{ij}$ we have
\begin{equation}
R = 2 \L_n K  + K^2 + \tr(K^2). 
\end{equation}
Here, in full,  $\tr(K^2) = (K^2)_{ij}\,\delta^{ij} = K_{ij} \,\delta^{ik}\,\delta^{jl} \,K_{kl}$, though we can and shall often simplify notation by writing  $\tr(K^2) = K_{ij} \, K_{ij}$, implying contraction on the repeated down indices.

\subsection{Einstein tensor }
Using $G_{ab} = R_{ab} - {1\over2} R g_{ab}$ we find
\begin{equation}
\label{E:Gnn}
G_{nn} =  {1\over2} \left(  K^2-\tr(K^2) \right),
\end{equation}
\begin{equation}
\label{E:Gni}
G_{ni} =  K_{ij,j}- K_{,i}, 
\end{equation}
and the somewhat messier result that
\begin{equation}
\label{E:Gij}
G_{ij} =  \L_n K_{ij}  + K K_{ij} - 2(K^2)_{ij} -  \left(\L_n K  + {1\over2}K^2 + {1\over2} \tr(K^2)\right) \delta_{ij}.
\end{equation}

\clearpage
Equivalently (and more explicitly)
\begin{equation}
\label{E:Gij2}
G_{ij} =  n^a \partial_a  K_{ij}  + K K_{ij}  + \partial_{[i} v_{k]} K_{kj} + K_{ik} \partial_{[j} v_{k]}
-  \left(n^a \partial_a K  + {1\over2}K^2 + {1\over2} \tr(K^2)\right) \delta_{ij}.
\end{equation}

\section{Warp-field stress-energy tensor}\label{stress}
In this section, we use the standard Einstein equations $G_{ab} = 8\pi T_{ab}$ and analyse the density, flux, and spatial components of the stress-energy. 

\subsection{Density} 
In the Eulerian orthonormal basis, using  (\ref{E:Gnn}), for the density we have:
\begin{equation}
\label{E:rho}
\rho = {G_{nn}\over 8\pi} = {1\over16\pi} \left(  K^2-\tr(K^2) \right).
\end{equation}
(Everyone agrees with this.  This is the quantity that  Lentz~\cite{Lentz:2020} and Fell--Heisenberg~\cite{Fell:2021}  eventually try to force to be positive. This is the only quantity Bobrick--Martire~\cite{Bobrick:2021} explicitly calculate.) 
\def\cof{{\mathrm{cof}}}
Now it is a mathematical identity that for any $3\times 3$ matrix
\begin{equation}
K^2-\tr(K^2)  = 2 \tr\left( \cof( K_{ij} )\right). 
\end{equation}
That is,  $K^2-\tr(K^2)$ is twice the trace of the cofactor matrix of the $3\times 3$ matrix $K_{ij}$. This is most easily established by diagonalizing the  $3\times 3$ matrix. So in (3+1) dimensions the Eulerian energy density is intimately related to the cofactor matrix of the extrinsic curvature. 
This is the ``hidden geometric structure'' alluded to by Fell--Heisenberg in reference~\cite{Fell:2021}, a curiosity which is unfortunately less useful than one might hope.

In the present context we have the fully explicit result 
\begin{equation}
\rho = {1\over16\pi} \left(  (v_{i,i})^2 -v_{(i,j)} \,v_{(i,j)}  \right).
\end{equation}
It is straightforward to rewrite this as a spatial divergence plus a negative semi-definite contribution:
\begin{equation}
\rho =
{1\over16\pi} \left\{ \partial_i ( v_i \,v_{j,j} - v_j \,v_{i,j})  - v_{[i,j]} v_{[i,j]}  \right\}.
\end{equation}
In terms of the vorticity $\omega_i = \epsilon_{ijk} \, v_{[j,k]}$ this becomes
\begin{equation}
\label{E:divergence}
\rho =
{1\over16\pi} \left\{ \partial_i ( v_i \,v_{j,j} - v_j \,v_{i,j})  - {1\over2} \, (\omega_i \, \omega_i) \right\},
\end{equation}
or even
\begin{equation}
\label{E:divergence-x}
\rho =
{1\over16\pi} \left\{ \div \{ \vec v\,K - (\vec v \cdot\grad)  \vec v\}  - {1\over2} \, (\vec\omega \cdot \vec\omega) \right\}.
\end{equation}
\enlargethispage{40pt}
That is, for the generic warp field, the Eulerian energy density is always the sum of a 3-divergence plus a quantity that is negative semi-definite. 
We will have occasion to repeatedly use this result in subsequent discussion. 

\subsection{Flux} 
For the flux, in the Eulerian orthonormal basis, using  (\ref{E:Gni}) we have:
\begin{equation}
f_i = {G_{ni}\over 8\pi} = {1\over8\pi} \left(K_{ij,j}- K_{,i} \right) 
= {1\over 16\pi} \left( v_{i,jj} - v_{j,ji} \right) 
= {1\over 16\pi} \left( \grad^2 v_i - \grad_i (\div \vec{v})  \right)
\end{equation}
That is
\begin{equation}
\label{E:flux}
f_i = {1\over 16\pi} \left( \curl (\curl \vec v) \right)_i
\end{equation}
(Everyone agrees with this. Note that  in the Lentz/Fell-Heisenberg~\cite{Lentz:2020,Fell:2021}  framework $\vec v = \grad\Phi$ is a gradient, so this co-moving Eulerian flux is identically zero.) 

\subsection{Spatial stresses} 
Finally, using  (\ref{E:Gij}), the spatial stresses are given by the $3\times3$ matrix:
 \begin{equation}
 \label{E:stress}
T_{ij} = {G_{ij}\over 8\pi} =
 {1\over8\pi} \left( \L_n K_{ij}  + K K_{ij} - 2(K^2)_{ij} -  \left(\L_n K  + {1\over2}K^2 + {1\over2} \tr(K^2)\right) \delta_{ij}\right).
\end{equation}
So for the average pressure, which we define to be $\bar p = {1\over 3} T_{ij} \, \delta^{ij}$ , and noting that
\begin{equation}
\left( \L_n K_{ij}  + K K_{ij} - 2(K^2)_{ij} \right) \delta^{ij} 
= \L_n K + K^2 -2 \tr(K^2) - K_{ij} \; \L_n \delta^{ij} = \L_n K + K^2,
\end{equation}
we find
\begin{equation}
\bar p = {1\over 3}\,{T_{ij} \,\delta^{ij}}  ={1\over24\pi} \left( -2 \L_n K -{1\over2} K^2 - {3\over2} \tr(K^2)\right).
\end{equation}
It is furthermore useful to note that
\begin{equation}
\label{E:sec-1}
\rho+3\bar p  = -{1\over4\pi}  \left(  \L_n K  + \tr(K^2) \right),
\end{equation}
and that
\begin{equation}
\label{E:nec-1}
\rho+\bar p = {1\over24\pi} \left( -2 \L_n K + K^2 - 3 \tr(K^2) \right).
\end{equation}
These quantities will soon be seen to be useful when investigating violations of the SEC and NEC, respectively. 

It is sometimes useful to also note that 
\begin{equation}
\nabla_a n^a = \partial_a n^a = K,
\end{equation}
 and so write
\begin{equation}
\L_n K = n^a\, \nabla_a K = \nabla_a ( K n^a) - K^2,
\end{equation}
whence
\begin{equation}
\bar p = -{1\over24\pi} \left(  2 \nabla_a ( K n^a) -{3\over2} K^2 + {3\over2} \tr(K^2) \right). 
\end{equation}
Equivalently
\begin{equation}
\bar p= \rho  -{1\over12\pi} \;  \nabla_a ( K n^a). 
\end{equation}
That is, for the generic warp field, the average spatial pressure equals the energy density plus  a 4-divergence. We will have occasion to repeatedly use this result in subsequent discussion. 

\enlargethispage{40pt}

For completeness we point out that the trace-free part of the $3\times3$ spatial stress tensor,
$ (T_{ij})^\mathrm{tf} =T_{ij} - {1\over 3}\;\bar p\; \delta_{ij}$,  can be computed in terms of the trace-free part of the extrinsic curvature, $(K_{ij})^\mathrm{tf} = K_{ij} - {1\over 3}\; K\; \delta_{ij}$.

We find 
\begin{equation}
 \label{E:stress-tf1}
(T_{ij})^\mathrm{tf}  =
 {1\over8\pi} \left\{ n^a \partial_a (K_{ij})^\mathrm{tf}  + K (K_{ij})^\mathrm{tf}
 + \partial_{[i} v_{k]} (K_{kj})^\mathrm{tf}  + (K_{ik})^\mathrm{tf}  \partial_{[j} v_{k]}
  \right\}.
 \end{equation}
That is
\begin{equation}
 \label{E:stress-tf2}
(T_{ij})^\mathrm{tf} = 
 {1\over8\pi} \left\{ \partial_a  [n^a (K_{ij})^\mathrm{tf}]
 + \partial_{[i} v_{k]} (K_{kj})^\mathrm{tf}  + (K_{ik})^\mathrm{tf}  \partial_{[j} v_{k]}
  \right\}.
 \end{equation}

Overall, we see that the generic Nat\'ario line element~(\ref{E:line}) is sufficiently simple to permit detailed and explicit calculations, while being sufficiently general to encompass almost all physically interesting warp drive spacetimes.

\section{Energy conditions}\label{S:ECIntro}
The (classical point-wise) energy conditions are constraints one places on the stress-energy as a way of keeping unusual physics somewhat under control. The energy conditions  all correspond, in some sense, to demanding that for some \emph{class} of observers the energy density be non-negative. 

\clearpage
Standard definitions are~\cite{ADM2,Wald,Poisson,Hawking-Ellis,Book,Carroll,Centenary,Curiel:2014,Kontou,Maeda:2018}:

\paragraph{NEC:} For all null vectors $k^a$ we demand $T_{ab} \; k^a k^b \geq 0$. 

\paragraph{WEC:} For all timelike vectors $V^a$ we demand $T_{ab} \; V^a V^b \geq 0$. 

\paragraph{SEC:} For all timelike vectors $V^a$ we demand $(T_{ab} - {1\over2} T g_{ab}) V^a V^b \geq 0$. 

\paragraph{DEC:} For all future-pointing timelike vectors $V^a$ and $W^a$ we demand $T_{ab} V^a W^b \geq 0$. \\
\null\qquad\qquad (The DEC in particular has a number of equivalent formulations~\cite{Kontou}.)

 The standard linkages between energy conditions are~\cite{ADM2,Wald,Poisson,Hawking-Ellis,Book,Carroll,Centenary,Curiel:2014,Kontou,Maeda:2018}:
\begin{equation}
\DEC \implies \WEC \implies \NEC
\end{equation}
\begin{equation}
\SEC \implies \NEC
\end{equation}
But 
\begin{equation}
\WEC\;\; \not\!\!\!\Longleftrightarrow\; \SEC
\end{equation}
Despite claims made in early versions of~\cite{Fell:2021}, the WEC is not the weakest of the energy conditions.

Some examples of the subtleties involved are:
\vspace{-10pt}
\begin{itemize}
\itemsep-3pt
\item 
A positive cosmological constant (positive vacuum energy density) satisfies the NEC and WEC, but not the SEC.
\item 
A negative cosmological constant (negative vacuum energy density) satisfies the NEC and SEC, but not the WEC.
\item 
Massive scalar fields, (classical, minimally coupled, with positive kinetic energy, and positive mass squared so that they are not tachyonic), satisfy the NEC and WEC but can violate the SEC.
For example the standard scalar inflaton field of cosmological inflation satisfies the NEC and WEC but violates the SEC.
\item
Classical Maxwell electromagnetism satisfies all the usual energy conditions.
\end{itemize}
\vspace{-10pt}
These examples suggest some caution when interpreting the physical significance of the energy conditions~\cite{twilight,cosmo99,epoch,science,galaxy,cosmodynamics,milestones,zee, Kontou:2020,trace}.

There are indications that NEC and WEC can be violated on microscopic (quantum) scales~\cite{FEC1,FEC2,Martin-Moruno:2013-quantum1, Martin-Moruno:2015-quantum2, gvp1,gvp2,gvp3,gvp4,gvp5,Fewster:2002,Fewster:2010,Fewster:2012, scale-anomalies, Flanagan:1996,Ford:1994,Ford:1995,Ford:2003,Ford:2005,Wald:anec}, 
though they seem to be satisfied by (reasonable) matter on macroscopic scales. In contrast the SEC  appears to be observationally violated on the largest cosmological scales~\cite{cosmo99,epoch,science,galaxy,cosmodynamics,milestones,Kontou:2020}. 

While for the purposes of this paper we will be focussing on applying these point-wise  energy conditions specifically to warp drive spacetimes, it should be noted that these energy conditions also have direct applications to singularity theorems, positive-mass theorems, traversable wormholes~\cite{Visser:2003,Hochberg:1998a,Hochberg:1998b,Hochberg:1998c,Barcelo:2000,Barcelo:brane,Kar:2004,Roman:1983,Roman:wormhole,Lobo:2016}, topological censorship, and chronology protection~\cite{hawking60,Visser:1992a,Visser:1992b,Friedman:2008,do-not-mess}.

Other less commonly used energy conditions are:
\begin{itemize}
\item 
TEC --- trace energy condition: $\rho-3p\geq 0$. Mainly of historic interest~\cite{twilight,trace}. \\
Believed to be \emph{violated} deep in the cores of neutron stars. Definitely violated by (hypothetical) ``stiff matter'' $\rho=p$, where (speed of sound) = (speed of light). 
\item
FEC --- flux energy condition: for all  timelike observers, $V^a$, the flux 
$F^a = T^{ab}V_b$\\ is either timelike or null~\cite{FEC1,FEC2}.\\
 (FEC is a weakening of DEC; that is, DEC $\implies$ FEC.)
 \enlargethispage{40pt}
\item
Averaged energy conditions (typically averaged along timelike or null geodesics) are weaker than their corresponding point-wise counterparts. The most useful of the averaged energy conditions is typically the ANEC (averaged null energy condition) where one averages the NEC along timelike geodesics \emph{using an affine parameterization}~\cite{Book,scale-anomalies, Flanagan:1996, Ford:1994,Ford:1995,Ford:2003,Ford:2005,Wald:anec}. 
\end{itemize}

As a cautionary example regarding application of the energy conditions, note that all three of the recent Lentz~\cite{Lentz:2020},  Bobrick--Martire~\cite{Bobrick:2021}, and Fell--Heisenberg~\cite{Fell:2021} articles merely assert the existence of \emph{one} sub-class of timelike observers for which the energy density is positive. 
This is \emph{not} enough to show that the WEC is satisfied.
To explicitly see a specific example of this behaviour, let us work in an orthonormal basis. Take $\rho_0>0$ and $\Gamma>1$, and consider:
\begin{equation}
T_{\hat a\hat b} = \rho_0 \left[\begin{array}{c|ccc}
1&0&0&0\\ \hline 0 &-\Gamma^2&0&0\\0&0&-\Gamma^2&0\\0&0&0&-\Gamma^2\\
\end{array}\right]_{\hat a\hat b}
\end{equation}
Then in the natural rest frame $V^{\hat a}=(1;0,0,0)^{\hat a}$ we have $\rho = T_{\hat a\hat b} V^{\hat a} V^{\hat b} = \rho_0 >0$.
But a moving observer, with 4-velocity $V^{\hat a}=\gamma(1; v \; n^i)^{\hat a}$, where $n^i$ is any unit spatial 3-vector,  will see an energy density
\begin{equation}
\rho = T_{\hat a\hat b} \; V^{\hat a} V^{\hat b} = \rho_0 \gamma^2 (1 - \Gamma^2 v^2). 
\end{equation}
For a sufficiently rapidly moving but still subluminal observer, with $|v| > 1/\Gamma$, the energy density will be seen to be negative.
So in this example the WEC is \emph{violated}. What Lentz~\cite{Lentz:2020},  and Bobrick--Martire~\cite{Bobrick:2021}, and Fell--Heisenberg~\cite{Fell:2021} \emph{should} be doing is to also calculate all of the stress components $T_{ij}$. It is not enough for them to just focus on $T_{nn}$ and $T_{ni}$. 

One way of proceeding is to assume the stress-energy tensor is of Hawking--Ellis type~I, see references~\cite{Hawking-Ellis,Book,Curiel:2014,FEC2,Martin-Moruno:2018core,Martin-Moruno:2017rainich,Martin-Moruno:2018-typeIII,Martin-Moruno:2019-ugly,Martin-Moruno:2021-back-reaction}, and work with the Lorentz-invariant eigenvalues $\{\rho,p_i\}$, the  Lorentz-invariant eigen-energy-density and principal pressures, respectively. 

Then it is a standard result that in terms of the Lorentz-invariant eigenvalues of a type~I stress-energy tensor one has the two-way implications~\cite{Hawking-Ellis,Book,Curiel:2014,FEC2,Martin-Moruno:2018core,Martin-Moruno:2017rainich,Martin-Moruno:2018-typeIII,Martin-Moruno:2019-ugly,Martin-Moruno:2021-back-reaction} 
\enlargethispage{20pt}
\begin{eqnarray}
\NEC &\iff& \rho+ p_i \geq 0.
\\
\WEC  &\iff& \rho+ p_i \geq 0 \quad \&\quad  \rho >0.
\\
\SEC  &\iff& \rho+ p_i \geq 0  \quad \&\quad  \rho+ \textstyle{\sum_i p_i}  >0.
\\
\DEC &\iff& |p_i| \leq \rho.
\end{eqnarray}\enlargethispage{20pt}
Unfortunately, we do not know \emph{a priori} whether or not  the stress-energy for the warp drive is Hawking--Ellis type I in general, nor do the Eulerian observers typically diagonalize the stress-energy. 
So one has to be a little more indirect. 
(A limited result is this: For the zero-vorticity warp drives, because the Eulerian observers see zero flux $\vec f=0$, the stress-energy is block diagonal, and so of Hawking--Ellis type I. In particular, an even more limited and precise result is this: For a zero-vorticity warp drive, since the stress-energy is Hawking--Ellis type I we explicitly have
		\begin{eqnarray}
			T_{ab} \; V^a V^b &=& \gamma^2 \left(\rho + \sum_i p_i \beta_i^2\right) =
			\gamma^2 \left(\rho(1-\beta^2) + \sum_i [\rho+ p_i ]\beta_i^2\right) 
			\nonumber\\ &=&
			\rho + \gamma^2 \left( \sum_i [\rho+ p_i ]\beta_i^2\right).
		\end{eqnarray}
		Thus, by considering all possible physical values of the $\beta_i$ (including relativistic values), we see that non-negative Eulerian density \emph{plus} the NEC implies the WEC --- so for checking the WEC in this situation, checking $\rho\geq0$ is manifestly insufficient, one additionally needs to check the NEC ($\rho+p_i\geq 0$) as well.)

In counterpoint, what the Eulerian observers do generically implement is a natural orthonormal frame in which 
\begin{equation}
T_{\hat a \hat b} = \left[\begin{array}{c|c} \rho & f_j \\ \hline  f_i & T_{ij} \end{array} \right] .
\end{equation}

\paragraph{NEC:} Let us first investigate what we can say about the NEC.  Let us take any two oppositely oriented null vectors $\ell_+^{\hat a} = (1,+\ell^i)^{\hat a}$, and $\ell_-^{\hat a} = (1,-\ell^i)^{\hat a}$, where $\ell^i$ is any arbitrary unit spatial 3-vector. Then the NEC would imply \emph{both}
\begin{equation}
T_{\hat a \hat b} \, \ell_+^{\hat a} \ell_\pm^{\hat b} =
T_{\hat a \hat b}  \, (1,+\ell^i)^{\hat a} (1,+\ell^j)^{\hat b} 
= \rho + 2 f_i \ell^i + T_{ij} \ell^i\ell^j \geq 0,
\end{equation}
\emph{and}
\begin{equation}
T_{\hat a \hat b} \, \ell_-^{\hat a} \ell_-^{\hat b} =
T_{\hat a \hat b}  \, (1,-\ell^i)^{\hat a} (1,-\ell^j)^{\hat b} 
= \rho - 2 f_i \ell^i + T_{ij} \ell^i\ell^j \geq 0.
\end{equation}

Averaging over these two equations, for any unit spatial 3-vector
\begin{equation}
\NEC \implies \rho + T_{ij} \ell^i \ell^j \geq 0.
\end{equation}
Note the implication is one-way; effectively one throws away all information contained in the flux $f^i$. In terms of the eigenvalues of the 3-stress we have\footnote{For zero-vorticity warp drives one has the stronger \emph{equivalence} that $\NEC \Longleftrightarrow \rho + \min\{ \lambda(T_{ij}) \} \geq 0$.}

\begin{equation}
\NEC \implies \rho + \min\{ \lambda(T_{ij}) \} \geq 0.
\end{equation}

Now pick a triad $\ell^i_A$ of three mutually orthogonal unit vectors. 
Then for each member of the triad the NEC implies
\begin{equation}
\rho + T_{ij} \, \ell^i_A \, \ell^j_A \geq 0.
\end{equation}
Now average over the three members of the triad
\begin{equation}
\rho + T_{ij} \left( {1\over 3}\ \sum_A \ell^i_A \ell^j_A \right) \geq  0.
\end{equation}
\enlargethispage{20pt}
Thence, noting that by construction $\sum_A \ell^i_A \ell^j_A= \delta^{ij}$, we see
\begin{equation}
\rho + T_{ij} \left( {1\over 3} \,\delta^{ij} \right) \geq 0.
\end{equation}

Defining, as usual, the average pressure as $\bar p = {1\over3} T_{ij}\,\delta^{ij}$, (this works even if the 3-stress $T_{ij}$ is not diagonal), we see that
\begin{equation}
\NEC \implies \rho + \bar p \geq  0.
\end{equation}

As long as the 3-stress is even slightly anisotropic we have $\min\{ \lambda(T_{ij}) \} \neq \max\{ \lambda(T_{ij}) \} $ so that we deduce the strict chain of inequalities
\begin{equation}
\min\{ \lambda(T_{ij}) \} < \bar p < \max\{ \lambda(T_{ij}) \}
\end{equation}
Thence in the presence of anisotropies we can make the slightly stronger strict inequality involving the average pressure $\bar p$ that
\begin{equation}
\NEC \implies \rho + \bar p > 0.
\end{equation}

The implication is again one-way.
Note also that this argument has nothing specific to do with warp drives, it is purely a statement about how the NEC implies an easily checked inequality that depends only on the existence of some orthonormal basis. For our current purposes the point is that it is relatively easy to pick a specific direction and calculate  $T_{ij}\; \ell^i \ell^j$, or to average over all directions and calculate $\bar p$.

\paragraph{WEC:} Similar things can be said about the WEC, although now one takes some $0 \leq \beta<1$ and considers two oppositely oriented timelike vectors $V^a_\pm = \gamma( 1,\pm \beta \ell^i)^a$. Then for any unit 3-vector $\ell^i$, after averaging over the two orientations:
\begin{equation}
\WEC \implies \rho + \beta^2\, T_{ij} \, \ell^i \ell^j \geq 0.
\end{equation}
Then averaging over a triad of unit vectors, for all $0 \leq \beta<1$ one has
\begin{equation}
\WEC \implies \rho + \beta^2 \, \bar p \geq 0.
\end{equation}
By considering $\beta=0$ and the limit $\beta\to 1$ one has
\begin{equation}
\WEC \implies \rho \geq 0 \quad \& \quad \rho +\bar p \geq 0.
\end{equation}
The implication is again one-way. And the point is that $ T_{ij} \, \ell^i \ell^j $ and $\bar p$ are relatively easy to calculate. 
As long as the 3-stress is even slightly anisotropic, we can make the slightly stronger statement involving the average pressure $\bar p$ that
\begin{equation}
\WEC \implies \rho \geq 0 \quad \& \quad \rho +\bar p > 0.
\end{equation}

\paragraph{SEC:} The SEC can be phrased in terms of the so-called ``trace-reversed'' stress-energy tensor $T_{ab}-{1\over 2} T g_{ab}$, as the condition $(T_{ab}-{1\over 2} T g_{ab})V^a V^b \geq 0$, which in turn is equivalent to enforcing $T_{ab} V^a V^b \geq {1\over2}(\rho-3\bar p)$.  Then the SEC implies
\begin{equation}
 \rho + \beta^2\, T_{ij} \, \ell^i \ell^j \geq (1-\beta^2) \; {1\over2}\; (\rho-3\bar p).
\end{equation}
This can be rearranged to
\begin{equation}
\SEC \implies  (1+\beta^2) \rho + {3}(1-\beta^2)\bar p + 2\beta^2\, T_{ij} \, \ell^i \ell^j \geq 0.
\end{equation}
Averaging over the unit directions $\ell^i$, for all $0 \leq \beta<1$ one has
\begin{equation}
\SEC \implies (1+\beta^2) \rho + (3-\beta^2)\bar p \geq 0.
\end{equation}
By considering $\beta=0$ and the limit $\beta\to 1$ one has
\begin{equation}
\SEC \implies \rho + 3\bar p \geq 0 \quad \& \quad \rho + \bar p \geq 0.
\end{equation}
The implication is again one-way. And the point is that $ T_{ij} \, \ell^i \ell^j $ and $\bar p$ are relatively easy to calculate. 
As long as the 3-stress is even slightly anisotropic, we can make the slightly stronger statement involving the average pressure $\bar p$ that
\begin{equation}
\SEC \implies \rho + 3\bar p > 0 \quad \& \quad \rho + \bar p > 0.
\end{equation}

\enlargethispage{10pt}
\paragraph{DEC:} One version of the DEC, as reported by Hawking and Ellis~\cite[page 91]{Hawking-Ellis}, is formulated as the requirement that in any orthonormal frame the energy density dominates all the other components of the stress-energy tensor:
\begin{equation}
| T^{\hat a \hat b} | \leq T^{\hat t\hat t}.
\end{equation}
In our language this would be
\begin{equation}
| f_i | \leq \rho \quad \& \quad |T_{ij}| \leq \rho.
\end{equation}
But in particular this implies 
\begin{equation}
|\bar p| =  \left| {1\over3} \sum_i T_{ii} \right| \leq {1\over3} \sum_i \left| T_{ii} \right| \leq \rho.
\end{equation}
That is
\begin{equation}
\DEC \quad\implies\quad |\bar p| \leq \rho.
\end{equation}
The implication is one-way, but the inequality is particularly clean and easy to work with.
As long as the 3-stress is even slightly anisotropic, we can make the slightly stronger statement involving the average pressure $\bar p$ that
\begin{equation}
\DEC \quad\implies\quad |\bar p| < \rho.
\end{equation}

\section{Timelike and null convergence conditions}\label{S:CC}

In applications the various energy conditions are, using the Einstein equations,  typically immediately converted into purely geometrical convergence conditions such as the null convergence condition (NCC) and the timelike convergence condition (TCC)~\cite{Book,Hawking-Ellis,Curiel:2014,Kontou}. Within standard general relativity the NCC is equivalent to the NEC, and the TCC is equivalent to the SEC. 

\paragraph{NCC:} The NCC is the statement that for all null vectors $R_{ab} \,\ell^a\ell^b \geq 0$.

\paragraph{TCC:} The TCC is the statement that for all timelike vectors $R_{ab}\, V^aV^b \geq 0$.

If one wishes to step outside the framework of standard general relativity then this adds a new level of speculative physics to the mix, and then the distinction between geometrical convergence conditions and dynamical energy conditions might become important. However it should be emphasized that in many (but not all) modified theories of gravity the equations of motion can be rearranged into the form 
\begin{equation}
\hbox{(Einstein tensor) = (some ``effective'' stress-energy tensor)}. 
\end{equation}
Whenever this can be done, statements about the usual energy conditions in Einstein gravity can be carried over to statements about ``effective'' energy conditions in modified gravity.

\section{Violation of the energy conditions}\label{S:ECs}

Now consider the energy conditions in the warp drive spacetimes. We shall provide a number of results, ultimately demonstrating the generic violation  of the NEC (thereby automatically implying violations of the WEC, SEC and DEC)

\subsection{Alcubierre warp drive}
The key defining characteristic of the Alcubierre warp drive is that $v_i(x,y,z,t)$ is always pointing in some fixed direction, which can without loss of generality be taken to be the $z$-direction  $\hat z_i=(0,0,1)_i $. That is, take
\begin{equation}
v_i(x,y,z,t) = v(x,y,z,t) \; \hat z_i.
\end{equation}
(This is slightly more general than what Alcubierre actually did~\cite{Alcubierre:1994}; but it is the best way of summarizing the key aspects of the physics.) 
Then the extrinsic curvature is simply
\begin{equation}
K_{ij} = \partial_{(i} v \; \hat z_{j)} = \left[ \begin{array}{cc|c} 
0&0&{1\over2}{\partial_x v} \\ 0&0&{1\over2} {\partial_y v} \\ \hline
{1\over2}{\partial_x v} &{1\over2}{\partial_x v}  & \partial_z v 
\end{array}\right].
\end{equation}
Consequently
\begin{equation}
K = \hat z^i \; \partial_i v = \partial_z v; 
\end{equation}
while
\begin{equation}
(K^2)_{ij} = {1\over2} K K_{ij} + {1\over4} (\partial v)^2 \hat z_i \hat z_j 
+ {1\over4} \partial_i v \partial_j v;
\end{equation}
and
\begin{equation}
\tr(K^2) = {1\over2} K^2 + {1\over2}  (\partial v)^2 =  {1\over2}(\partial_x v)^2 +  {1\over2}(\partial_y v)^2 +  (\partial_z v)^2.
\end{equation}
Then
\begin{equation}
\rho =  {1\over16\pi} \left(  K^2-\tr(K^2) \right) = 
-{1\over32\pi} \left( (\partial_x v)^2 + (\partial_y v)^2 \right)
 \leq 0.
\end{equation}
But to 
have a non-trivial warp bubble, we must
avoid the situation where one has an everywhere flat Minkowski space. So  we need $  \left( (\partial_x v)^2 + (\partial_y v)^2 \right) >0$ somewhere in the spacetime, and so $\rho < 0$ at those particular locations, and is at best zero everywhere else. 
Therefore the Alcubierre warp bubble definitely violates the WEC. 

In his original article~\cite{Alcubierre:1994} Alcubierre also claims (without proof): 
``In a similar way one can show that the strong energy condition is also violated.''
However in a more recent article by Alcubierre and Lobo~\cite{Alcubierre:2017}, they explicitly prove the stronger result that the Alcubierre warp drive violates the NEC. 
(So SEC, WEC, and DEC are all definitely violated, because NEC is.)

Their proof is slightly tricky and depends (for a warp bubble moving in the $z$ direction) on the identity $G_{zz}= 3 G_{nn}<0$.
This identity is established by explicit computation, there does not seem to be an obvious geometrical reason for it.  From this geometrical identity we have $T_{zz} = 3 \rho < 0$ in the comoving frame. Hence $\rho + T_{zz} = 4 \rho <0$ and the NEC is violated in all Alcubierre warp drives.

\subsection{Nat\'ario zero-expansion warp drive}
The key defining characteristic of the Nat\'ario zero-expansion warp drive is that the flow $v_i(x,y,z,t)$ is zero divergence:  $\div \vec v=0$. (This is slightly more general than what Nat\'ario actually did~\cite{Natario:2001}; but it is the best way of summarizing the relevant physics.) 
Then $ K = 0$ and from (\ref{E:rho}) we have:
\begin{equation}
\rho =  {1\over16\pi} \left(  K^2-\tr(K^2) \right) = -{1\over16\pi} \tr(K^2) \leq0.
\end{equation}
But to avoid a trivial warp drive the extrinsic curvature $K_{ij}$ must be nonzero somewhere, 
and wherever $K_{ij}\neq 0$ we have $\rho<0$.
So the Nat\'ario zero-expansion warp drive definitely violates the WEC~\cite{Natario:2001}. 

But, now extending Nat\'ario's argument from the second half of reference~\cite{Natario:2001},  we note that in this situation, from equation (\ref{E:sec-1}), we also have
\begin{equation}
\rho+3\bar p = -{1\over4\pi}   \tr(K^2) \leq 0.
\end{equation}
And the same non-triviality argument as immediately above now shows that any Nat\'ario zero-expansion warp drive also violates the SEC. 

Furthermore, again extending Nat\'ario's argument from the second half of reference~\cite{Natario:2001}, now using equation (\ref{E:nec-1}) we note
\begin{equation}
\rho+\bar p = -{1\over8\pi}   \tr(K^2) \leq 0.
\end{equation}
And the same non-triviality argument again shows that any Nat\'ario zero-expansion warp drive also violates the NEC.

\subsection{Zero-vorticity warp drive}
The key defining characteristic of the Lentz/Fell--Heisenberg zero-vorticity warp drive is that the flow $v_i(x,y,z,t)$ is taken to be a gradient  $v_i(x,y,z,t) = \partial_i \Phi(x,y,z,t)$~\cite{Lentz:2020,Fell:2021}. (This is slightly more general than what Lentz and Fell--Heisenberg actually did; but it is the best way of summarizing the relevant physics.) 

Then
\begin{equation}
K_{ij} = \Phi_{,ij}, \qquad K = \grad^2 \Phi.
\end{equation}
From (\ref{E:rho}) the co-moving energy density is 
\begin{equation}
\rho =  {1\over16\pi} \left(  K^2-\tr(K^2) \right) 
={1\over16\pi} \left(  (\grad^2\Phi)^2-  \Phi_{,ij} \Phi_{,ij} \right).
\end{equation}
This no longer \emph{obviously}  violates the WEC, at least not as seen by Eulerian observers.
In view of equation (\ref{E:divergence}), the energy density is now a pure 3-divergence
\begin{equation}
\label{E:divergence2a}
\rho =
{1\over16\pi} \left\{ \partial_i ( \Phi_{,i} \Phi_{,jj} - \Phi_{,j} \,\Phi_{,ij})  \right\}.
\end{equation}
In this class of zero-vorticity warp drives the comoving flux is identically zero, $f^i=0$.\\
Unfortunately, the spatial parts of the stress-energy $T_{ij}$ are still a bit of a mess, they do not really simplify appreciably. (And one really needs to know something about the spatial stresses $T_{ij}$ in order to say anything more precise about the energy conditions.)
Fortunately, the average pressure is reasonably tractable. From equation (\ref{E:nec-1}), 
\begin{equation}
\label{E:nec-1b}
\rho+\bar p 
= {1\over24\pi} \left( -2 \L_n K + K^2 - 3 \tr(K^2) \right)
={1\over24\pi} \left( -2 \L_n \grad^2\Phi 
+ (\grad^2\Phi)^2 - 
3 \Phi_{,ij} \; \Phi_{,ij} \right).
\end{equation}

Now, given that we want our warp field to be localizable, to not spread over and affect the whole spacetime, it is natural to make the assumption that the flow field $v_i(x,y,z,t) = \partial_i\Phi(x,y,z,t)$ 
obeys some sort of fall-off conditions at spatial infinity. 
	
It is certainly more than sufficient if $J_i = ( \Phi_{,i} \Phi_{,jj} - \Phi_{,j} \,\Phi_{,ij}) = o(r^{-2})$ for large distances, which is in turn certainly true if $\Phi_i = o(r^{-1/2})$, which is in turn certainly true if the ADM mass is zero. 
Vanishing of the ADM mass is certainly the case for Alcubierre's warp drive, and also for Natario's zero-expansion warp drive, and is a first crude approximation for well-localized warp fields.  (We will subsequently significantly weaken this condition.) 
If this is the case, then it follows immediately that

\begin{equation}
	\int_{\mathbb{R}^3} \rho \; \d^3 x = 0.
\end{equation} 

But then, if the warp field has positive energy density anywhere, it must have negative energy density somewhere else. This implies violations of the WEC somewhere on each spatial slice.
We shall subsequently be more specific by invoking stronger arguments using much weaker falloff conditions. 

Unfortunately, as we have previously argued, non-negativity of the Eulerian energy density, $\rho\geq0$, is insufficient to establish the WEC for a zero-vorticity warp drive; one needs to check the NEC as well.
The spatial parts of the stress-energy $T_{ij}$ are still a bit of a mess, they do not really simplify appreciably. (And one really needs to know something about the spatial stresses $T_{ij}$ in order to say anything more precise about the energy conditions.)
Fortunately, the average pressure is reasonably tractable. From equation (\ref{E:nec-1}), 
	\begin{equation}
		\label{E:nec-1b}
		\rho+\bar p 
		= {1\over24\pi} \left( -2 \L_n K + K^2 - 3 \tr(K^2) \right)
		={1\over24\pi} \left( -2 \L_n \grad^2\Phi 
		+ (\grad^2\Phi)^2 - 
		3 \Phi_{,ij} \; \Phi_{,ij} \right).
	\end{equation}

The best way of proceeding seems to be to adapt, modify, and extend a general argument that Nat\'ario applied to his general class of warp drive spacetimes~\cite{Natario:2001}. Let us do this now.

\subsection{Nat\'ario's generic warp drive}
\paragraph{WEC:}
We have already established that from equation~(\ref{E:divergence-x}) it follows in general that
\begin{equation}
	\label{E:divergence2b}
	\rho =
	{1\over16\pi} \left\{ \div \{ \vec v\,K - (\vec v \cdot\grad)  \vec v\}  - {1\over2} \, 
	(\vec\omega \cdot \vec\omega) \right\}.
\end{equation}

Hence, with the strong fall-off conditions already mentioned, namely that the warp field is well-localized, and that gradients of $v_i \to 0$ sufficiently rapidly at spatial infinity, an integration by parts implies 
\begin{equation}
	\int \rho \; \d^3 x = -{1\over 32\pi} \int (\vec\omega \cdot \vec\omega) \, \d^3 x \leq 0.
\end{equation}
Again,  if the warp field has positive energy density anywhere, it must have negative energy density somewhere else. This already implies violations of the WEC somewhere on each spatial slice.
(We shall establish stronger results using weaker hypotheses below.)

\paragraph{WEC-or-SEC:} 
To improve on the preceding result for the WEC, and side-step the need for strong asymptotic falloff conditions, note that in reference~\cite{Natario:2001} Nat\'ario presents a quite general argument to the effect that in \emph{any} generic warp spacetime, (Alcubierre, Nat\'ario zero-expansion, and, yes, it even applies to Lentz/Fell--Heisenberg zero-vorticity warp drives), there \emph{must} be violations of either the SEC or the WEC. (Or you have the trivial case of Minkowski space.) See his theorem 1.7.

\paragraph{SEC:} 
Nat\'ario's result can already be slightly improved in view of our comments above: In \emph{any} generic warp spacetime, (Alcubierre, Nat\'ario, and, yes, it even applies to Lentz/Fell--Heisenberg), there \emph{must} be violations of the SEC. (Or you have the trivial case of Minkowski space.)
We shall go into some detail in order to localise \emph{where} and \emph{when}  the SEC violations take place.

The key point is that Eulerian observers define a zero-vorticity congruence of timelike geodesics,
 that by construction \emph{cannot} have any focussing points. 
Now apply a variant of the Raychaudhuri equation (timelike focussing theorem)~\cite{Hawking-Ellis, Raychaudhuri,Dadhich:2005,Ehlers:2006,Borde:1987,vanVleck,Kar:2006,Abreu:2010}.

Explicit calculation has shown us, see equation  (\ref{E:sec-1}), that for \emph{any} warp drive spacetime:
\begin{equation}
\rho+3\bar p = -{1\over4\pi}  \left(  \L_n K  + \tr(K^2) \right).
\end{equation}
Thus if the SEC holds (implying $\rho+3\bar p \geq 0$) we have
\begin{equation}
  \L_n K  + \tr(K^2)  \leq 0.
\end{equation}
So
\begin{equation}
\L_n K  \leq - \tr(K^2).
\end{equation}
Now split the extrinsic curvature into trace-free part $K^\mathrm{tf}_{ij} = K_{ij} - {1\over 3} K \delta_{ij}$ and trace.
Then 
\begin{equation}
\tr(K^2) = \tr( \left[K^\mathrm{tf}_{ij}+ {1\over 3} K \delta_{ij}\right]^2) = \tr([K^\mathrm{tf}]^2)+ {1\over 3} K^2 
\geq  {1\over 3} K^2
\end{equation}
Consequently the SEC would imply \enlargethispage{20pt}
\begin{equation}
\L_n K  \leq -{1\over 3} K^2.
\end{equation}
So if $K<0$, then it will be even more negative in the future.
Similarly if $K>0$, then it must have been even more positive in the past.
Noting that, acting on scalars, $\L_n K  = V^a \partial_a K = \d K/\d\tau$, we see that the Eulerian observers satisfy
\begin{equation}
{\d K\over \d\tau}   \leq -{1\over 3} K^2.
\end{equation}

Therefore
\begin{equation}
- {1\over K^2} {\d K\over \d\tau}   \geq {1\over 3}.
\end{equation}
So
\begin{equation}
{\d K^{-1}\over \d\tau}   \geq {1\over 3}.
\end{equation}

Pick any Eulerian observer, and pick some point $\tau_0$ on that world-line. Integrating upwards, from $\tau=\tau_0$ to  some $\tau>\tau_0$ we have
\begin{equation}
K^{-1}(\tau)   \geq  K(\tau_0)^{-1} + {1\over 3}(\tau-\tau_0).
\end{equation}
So if $K(\tau_0)<0$, then there will be some finite proper time increment, less than $3/|K_0|$, at which $K^{-1}\to 0^-$ implying $K \to - \infty$. 

Integrating downwards, from $\tau=\tau_0$ to  some $\tau<\tau_0$ we have
\begin{equation}
K_0^{-1}  \geq   K(\tau)^{-1} + {1\over 3}|\tau-\tau_0|.
\end{equation}
That is, 
\begin{equation}
K^{-1}(\tau)   \leq  K_0^{-1} - {1\over 3}|\tau_0|.
\end{equation}
So if $K(\tau_0)>0$, then there will be some finite proper time decrement, less than $3/|K_0|$,  at which $K^{-1}\to 0^+$ implying $K \to + \infty$. 
\enlargethispage{20pt}

Either way, this is in contradiction to the fact that the Eulerian observers define a zero-vorticity congruence of timelike geodesics that by construction \emph{cannot} have any focussing points. 
So either $K\equiv 0$ identically, or whenever $K(\tau_0) \neq 0$ somewhere for any arbitrary Eulerian observer, the SEC fails somewhere in the interval  
\begin{equation}
\tau \in \left(\tau_0- {3\over|K(\tau_0)|},\;\tau_0+ {3\over|K(\tau_0)|}\right). 
\end{equation}
(Of course we could relax the global condition in the ADM-like split in the warp drive. However, this would imply either the formation of singularities in a finite time in the region of non-zero extrinsic curvature---meaning the warp bubble would encounter it; or we would have to save the situation by opting for CTCs~\cite{Everett:1995}, in many ways an even worse situation for physics~\cite{CPC,hawking60,Visser:1992a,Visser:1992b,Friedman:2008,do-not-mess}.)

Overall, this now gives us quite good control on  \emph{where} and \emph{when}  the SEC violations take place.
The special case where $K\equiv 0$ identically reduces to the Nat\'ario warp drive, for which we had already argued that the SEC fails (as long as the extrinsic curvature $K_{ij}$ is not identically zero everywhere in the spacetime). 

Note that this argument for SEC violations does not require any asymptotic conditions on the flow field at spatial infinity. 

\paragraph{NEC:} For the NEC we had already argued that the NEC requires $\rho+\bar p \geq 0$, and we have the explicit calculation (\ref{E:nec-1}):
\begin{equation}
\rho+\bar p = {1\over24\pi} \left( -2 \L_n K + K^2 - 3 \tr(K^2) \right).
\end{equation}
So the NEC would require
\begin{equation}
 2 \L_n K - K^2 + 3 \tr(K^2)  \leq 0.
\end{equation}
But, in terms of the trace-free part of extrinsic curvature, we have
\begin{equation}
\tr(K^2) = \tr([K^\mathrm{tf}]^2)+ {1\over 3} K^2. 
\end{equation}
So we can rewrite the NEC as
\begin{equation}
\L_n K \leq - {3\over2}  \tr([K^\mathrm{tf}]^2).  
\end{equation}
That is, using $\L_n K= \d K/\d \tau$, and  assuming the NEC:
\begin{equation}
	\label{E:dKdt}
{\d K\over \d\tau} \leq - {3\over2}  \tr([K^\mathrm{tf}]^2)\; \leq 0.
\end{equation}
Let us now take a moment to further explore equation \eqref{E:dKdt}. The $\tau$ here refers to the proper time along an Eulerian observer's trajectory. It represents, therefore, how a particular Eulerian observer will see the trace of the extrinsic curvature where they are located.
Given the zero-vorticity nature of the Eulerian observers' congruence, it is known that at each spacetime point one and only one Eulerian observer will be present.
We might, therefore, associate an Eulerian observer to each point on a constant-time spatial slice. 

Keeping this in mind, let us now follow a particular Eulerian observer who is initially sitting in a flat region far away from the warp bubble. Given the already mentioned falloff conditions for warp drives, we have that $K_{ij}$ tends to zero at spatial infinity. Therefore, as long as the observer is sufficiently distant from the warp bubble, $K_{ij}$ will be as close to zero as the particular fall-off conditions allow.

Without any loss of generality, we might pick such an observer to be in the region traversed by the warp drive as it moves from a certain origin to a specific destination --- also assumed to be sufficiently far away. In this particular scenario, the extrinsic curvature as seen by such an observer must start and end up with values which approach zero at spatial infinity --- since they start and finish 
in an almost flat space. 

Let us now return to equation \eqref{E:dKdt}. It tells us that, as long as the NEC is satisfied in the generic Natário warp drive spacetime, the trace of the extrinsic curvature can never increase. How does this apply to the particular Eulerian observer above?

In the situation just described, the trace as seen by the particular Eulerian observer will naturally start arbitrarily close to zero, since they are in an approximately flat space. In this way, once the warp drive passes, it can either decrease 
or stay the same. If is stays the same we have only two possible explanations: either a zero expansion warp drive passed by or no warp drive passed at all. 
Since we already proved the violation of the NEC for the zero expansion warp drive case separately, let us now focus on the other hypothesis, namely, that $K$ decreases.	
Now, again, two possibilities arise: $K$ becomes negative or $K$ becomes even closer to zero than it was in the asymptotically flat region.
Independently of the case, 
after the warp drive passes and proceeds sufficiently far away, the observer will return to 
an asymptotically flat region, meaning that $K$ increases as it returns to its original value.\footnote{Since we still have to assume that any contribution to $K_{ij}$ from the warp drive is sufficiently localized.}
 This, however, is not allowed if the NEC is always satisfied, as given by equation \eqref{E:dKdt}. This proves that the violation of the NEC is a necessary condition for the spacetime to 
``restore its 
asymptotics" after a warp drive passes through a particular point.

\begin{figure}
	\centering
	\includegraphics[width=.9\textwidth]{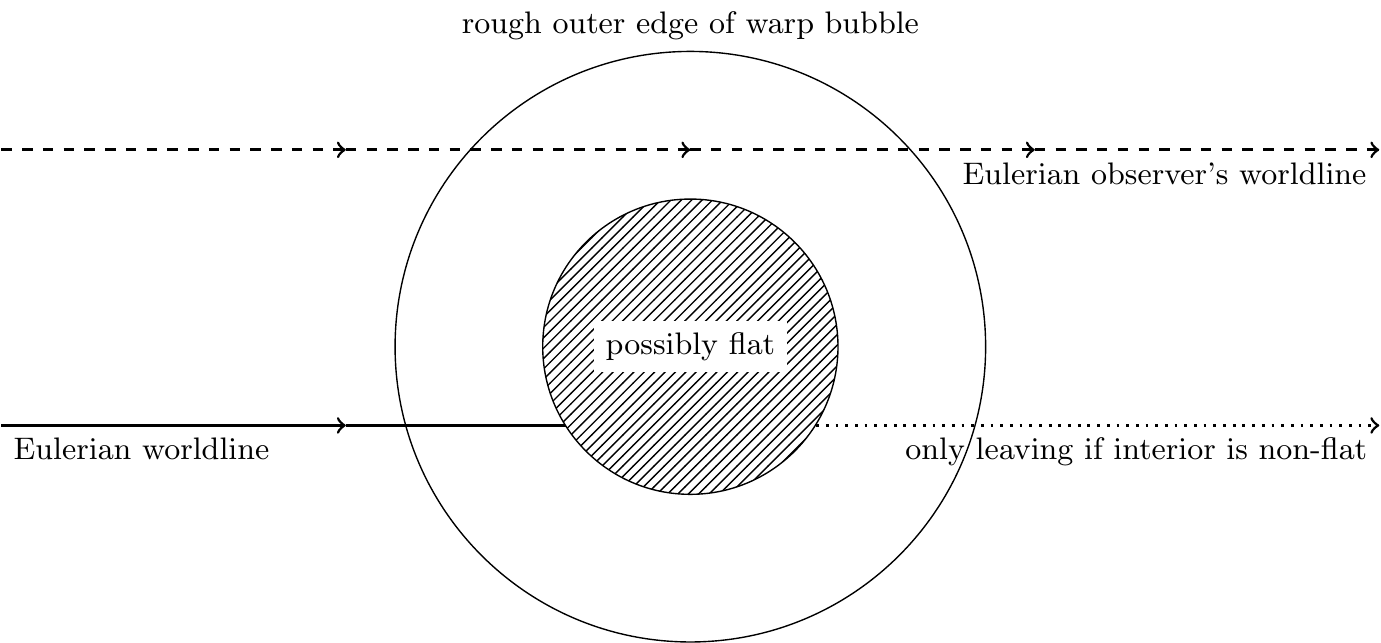}
	\caption{Illustration of the key step for the proof of the violation of the NEC, viewed from the perspective of the bubble's center. Eulerian observers as depicted here will be present in any warp drive. They will see a trace of the extrinsic curvature $K$ that is in violation of its monotonicity as predicted by the NEC. Eulerian observers entering a flat inner region will remain there as long as the bubble persists.}
	\label{fig:NEC}
\end{figure}

A delicate point must be discussed before this proof is called closed and complete: The question regarding what happens to such a geodesic Eulerian observer when a warp drive reaches them. Is the observer's trajectory disturbed by the warp drive? Is this geodesic dragged along so that it never leaves the bubble again? Gladly, all of these questions were already addressed in a paper by Pfenning and Ford \cite{PfenningFord}, where they analysed the effect of the passage of a constant velocity warp bubble on an Eulerian observer. 
Any observer passing through the bubble wall\footnote{We have to assume wall of finite size---but then again a wall of infinitesimal size would render the metric distributional and thus unsuitable for standard GR, anyhow.} but not reaching a flat interior will behave in the following way: As the front wall of the bubble reaches such Eulerian observers, their trajectory acquires a ``coordinate acceleration'' in the direction of the bubble, relative to observers at large distances.\footnote{The Eulerian observers are geodesic, so the 4-acceleration is identically zero.}
 For a while, this is followed by a movement with nearly constant speed, shown to be always smaller than the bubble's speed. These observers then finally decelerate, being left at rest as the bubble continues to go forward. In this way, while being displaced in space along the trajectory of the bubble, no momentum is transferred to these observers during the ``collision". Therefore, such Eulerian observers who were at rest before the arrival of the warp drive will finish as Eulerian observers at rest after the bubble passes. Another more recent study \cite{McMonigal:2012} extended these results to warp drives with non-constant velocities. An illustration of the Eulerian observers' worldlines in this argument for the violation of the NEC can be found in figure~\ref{fig:NEC}.\footnote{Very nice illustrations of this can also be found in \cite{GattermannBA2013}.}

Let us summarise this argument again: Eulerian observers initially at rest, which eventually ``collide'' with the bubble wall---without, however, entering possible flat regions inside the warp bubble---\emph{will exit} the bubble, and \emph{will again reenter} the asymptotically flat space behind the bubble. The monotonicity implied by equation~\eqref{E:dKdt} then implies a violation of the NEC somewhere along these Eulerian trajectories. It is, therefore, proven that any warp drive in the generic Natário form will inevitably violate the NEC as it moves through spacetime.

Note that this proof for NEC violations uses only the fact that warp drive contributions are sufficiently localized	
--- without any need to assume zero ADM mass or to invoke an integration by parts.

\clearpage
\section{Discussion and conclusions}\label{S:conclusions}
\vspace{-10pt}
\enlargethispage{40pt}
We have now demonstrated that all members of the general class of warp drives defined by Nat\'ario~\cite{Natario:2001} violate the NEC. This significantly extends previously known results. This argument applies, in particular, to all three of the Alcubierre~\cite{Alcubierre:1994}, Nat\'ario zero-expansion~\cite{Natario:2001}, and Lentz/Fell--Heisenberg zero-vorticity~\cite{Lentz:2020, Fell:2021} warp drives, and also (for slightly different reasons) to the model warp drives considered by Bobrick--Martire~\cite{Bobrick:2021}. Because the NEC is violated then so are the WEC, SEC, and DEC.

Consequently, insofar as one wishes to continue to entertain the possibility of warp drives as a real physical phenomenon, one has no choice but to face the violation of the energy conditions head on. 
Several possibilities arise: (i) modify the theory of gravity, (ii) modify the definition of warp drive, (iii)
modify the energy conditions, (iv) appeal to macroscopic quantum physics, (v) allow for singularities or CTCs (time travel). None of these options are particularly palatable. All of these options have serious draw-backs. Thus it is our melancholy duty to report that none of the recent claims of positive-mass physical warp drives survive careful inspection of the proffered arguments.

\appendices
\section{``Spherically symmetric'' warp drives}\label{S:spherical}

``Spherically symmetric'' warp drives date back to the original article by Alcubierre~\cite{Alcubierre:1994}. Specifically, let us reconsider the explicit line element~(\ref{E:A2}) and rephrase it as
\begin{equation} \label{E:line-ss}
     \d s^2 = - \d t^2 +\d x^2 +\d y^2 +  \Big( \d z -  v_*(t) \; f\left( r_s(x,y,z,t)\right) \d t\Big)^2,  
\end{equation}
 where we set
 \begin{equation}
r_s(x,y,z,t) = \sqrt{x^2+y^2+\left(z- z_*(t) \right)^2}\,, \qquad z_*(t)= \int v_*(t)\,\d t,
\end{equation}
and enforce both $ f(0)=1$ and $ f(\infty)=0$. This represents a spherically symmetric warp bubble, centred on the moving point $(0,0,z_*(t))$, with a shape function $f(r_s)$ that depends only on the Euclidean distance from the moving centre. We emphasize that while the warp bubble is spherically symmetric the spacetime is not --- there is certainly a preferred axis due to the direction of motion of the warp bubble. (Hence the warp bubble is spherically symmetric, but the spacetime is only ``spherically symmetric''.) 
At large spatial distances $\d s^2 \to  -  \d t^2 +\d x^2 +\d y^2+\d z^2$, the usual representation of Minkowski space, while at the centre of the warp drive 
\begin{equation} \label{E:line-ss2}
     \d s^2 \to  -  \d t^2 +\d x^2 + \d y^2 +  \left(\d z - v_*(t)  \d t\right)^2.  
\end{equation}
Consider now the coordinate transformation
\begin{equation}
\tilde t = t, \qquad \tilde x = x, \qquad \tilde y = y, \qquad \tilde z = z - z_*(t),
\end{equation}
which is designed to bring the warp bubble ``to rest''. Then
\begin{equation}
\d\tilde t = \d t, \qquad \d\tilde x = \d x, \qquad \d\tilde y = \d y, \qquad \d \tilde z = \d z - v_*(t)\, \d t ,
\end{equation}
and the line element becomes
\begin{equation} \label{E:line-ss3}
     \d s^2 = - \d \tilde t^2 +\d \tilde x^2 +\d \tilde y^2 
     +  \left( \d \tilde z + v_*(t) [1-f(r_s)]\, \d \tilde t\right)^2,  
\end{equation}
where now $r_s= \sqrt{\tilde x^2 +\tilde y^2 +\tilde z^2}$ is time independent. 
At the centre of the warp bubble we now have $\d s^2 \to  - \d \tilde  t^2 +\d \tilde x^2 +\d \tilde y^2+\d \tilde z^2$, the usual representation of Minkowski space, while at spatial infinity it is now the outside universe that is streaming by with velocity $-v_*(t)$. 

At spatial infinity the line element becomes
\begin{equation} \label{E:line-ss4}
     \d s^2 \to  - \d \tilde t^2 +\d \tilde x^2 +\d \tilde y^2 +  \left( \d \tilde z + v_*(t)\,  \d\tilde t\right)^2.  
\end{equation}
This is again Minkowski space, but now in ``moving coordinates''. 
Note the ``conservation'' of relative motion --- either spatial infinity is standing still and the bubble is moving, or the bubble is standing still and spatial infinity is moving~\cite{Alcubierre:1994,Natario:2001}. 

If one wishes to make the walls of the warp bubble thin, (but still of finite thickness), this is a \emph{choice}, not a \emph{necessity}, then one might pick
\begin{equation}
\label{E:smooth}
f(r_s) = \left\{ \begin{array}{ll}
1 & \quad r_s \leq r_{\text{inner}},\\
\hbox{smooth} & \quad r_s \in [r_{\text{inner}},r_{\text{outer}}],\\
0 & \quad r_s \geq r_{\text{outer}}.\\
\end{array}\right.
\end{equation}
If one makes this \emph{choice}, then one needs to enforce $r_{\text{outer}}> r_{\text{inner}}$ to keep the metric continuous. 
A discontinuity in the metric leads to delta functions in the Christoffel symbols, squares of delta functions in the Riemann tensor, and cubes of delta functions in the Bianchi identities. 
Even when working with the Israel--Lanczos--Sen thin-shell formalism (the junction condition formalism) one needs to keep the metric piecewise differentiable, $C^{1-}$, since then the Christoffel symbols at worst contain step functions and the Riemann tensor and Bianchi identities at worst contain delta functions.

\clearpage
The reason for being so explicit is that in reference~\cite{Bobrick:2021} the authors present a deeply flawed discussion of spherically symmetric warp drives.  In their implementation of spherically symmetric warp drives the warp bubble is certainly at rest, but in addition their choice of asymptotic boundary conditions also forces the warp bubble to be at rest with respect to spatial infinity. This defeats the whole purpose of a warp drive. 

Furthermore, these authors take the interior of the warp bubble to be a portion of flat Minkowski space, the wall of the warp bubble to be a non-negative density barotropic and isotropic fluid, and the exterior region to be a portion of Schwarzschild spacetime. They then attempt to apply the Tolman--Oppenheimer--Volkov (TOV) equation to the bubble wall. But under the set of assumptions they are imposing (zero pressure at the inner edge of the bubble wall, zero pressure at the outer edge of the bubble wall, non-negative energy density within the bubble wall, and a barotropic equation of state) the unique solution to the TOV is the trivial solution $p(r)=0=\rho(r)$, and the mass of the exterior region must be zero. Thus their specific model (as presented in~\cite{Bobrick:2021}) just reduces, globally, to flat Minkowski space --- it is not a warp drive. 

\enlargethispage{20pt}
These authors~\cite{Bobrick:2021} also assert that both Alcubierre~\cite{Alcubierre:1994} and Nat\'ario~\cite{Natario:2001} must \emph{necessarily} ``truncate'' their warp fields, forcing them to be exactly Minkowski space, at some finite distance from the centre of the warp bubble.  This is simply not an accurate reflection of what is actually done in those references~\cite{Alcubierre:1994,Natario:2001}. 
Specifically, in Alcubierre's original article~\cite[equation~(6)]{Alcubierre:1994} the concrete realization of the shape function $f(r_s)$ he chooses, see equation (\ref{E:smooth}), uses tanh functions and so does not have compact support. In this terminology, Alcubierre  sets $r_{\text{inner}}=0$ and $r_{\text{outer}}=\infty$, which is certainly not a truncation.
Unfortunately this terminology of ``truncation'' has been uncritically adopted by subsequent authors~\cite{Fell:2021}. 

The authors of~\cite{Bobrick:2021} also assert in passing, see their section~(5.2),  that the Alcubierre warp drive~\cite{Alcubierre:1994} does not satisfy the continuity equations. They also assert that the velocity of the Alcubierre warp bubble cannot be time dependent, claiming a violation of the conservation of momentum. These statements are both false, and seem to arise (at best) from adopting a naive Newtonian viewpoint.
What Alcubierre has actually done is to ``reverse engineer'' the warp drive --- once one writes down a suitable metric, (at least $C^{1-}$, piecewise differentiable), one simply calculates the Einstein tensor to find what the stress-energy is that would be required to support that spacetime geometry. 
The continuity equation is automatically enforced via the Bianchi identities, and there is no difficulty in making the velocity of the Alcubierre warp bubble time-dependent. (Reference~\cite{Fell:2021} seems to repeat this error in their discussion. Reference~\cite{Lentz:2020} is more careful in this regard, setting the 3-velocity of their model warp bubble constant for simplicity, but without making any claim as to necessity.)

Overall, while the idea of a ``spherically symmetric'' warp drive is certainly well-defined and useful,
the specific implementation of this notion in reference~\cite{Bobrick:2021} is not a successful one, and is not useful.

\section{Warp drives beyond the generic Natário framework}\label{A:B}

There are two somewhat more general warp drive spacetimes that do not fall into Nat\'ario's general 
classification~\cite{Natario:2001}, and are incompatible with the Alcubierre, Nat\'ario, and Lentz/Fell--Heisenberg warp drives.  These more general warp drives are obtained by either relaxing the condition that the lapse be unity, (so $N(x,y,z,t)\neq 1$), or relaxing the condition that the spatial slices be flat, (so $g_{ij} \neq \delta_{ij}$). See reference~\cite{Alcubierre:2017} which discusses the case $N\neq 1$, and reference \cite{VanDenBroeck:1999a} which permits the spatial slices to be conformally flat $g_{ij} = e^{2\theta(x,y,z,t)} \; \delta_{ij}$ rather than Riemann flat.

Unhelpfully, in appendices A.1 and A.2 of reference~\cite{Bobrick:2021} those authors incorrectly claim that
both of these more general spacetimes can, by a coordinate transformation, be brought into Alcubierre form. The simplest way to see that both these claims are wrong is to set the flow vector to zero. 
The claims made in appendices A.1 and A.2 of reference~\cite{Bobrick:2021} then reduce to the claims that the line elements
\begin{equation}
\d s^2 = - N(x,y,z,t)^2 \d t^2 + \d x^2 + \d y^2 +\d z^2,
\end{equation}
and 
\begin{equation}
\d s^2 = - \d t^2 + e^{2\theta(x,y,z,t)} \{\d x^2 + \d y^2 +\d z^2\},
\end{equation}
are actually Riemann flat. This is manifestly incorrect. What has gone wrong? The transformations these authors invoke are just not coordinate transformations. For a transformation $\d x^a \to \d \bar{x}^a = J^a{}_b \; \d x^b$ to actually be a coordinate transformation the critical requirement is that $J^a{}_{[b,c]} =0$.  (Additionally, one would want $\det(J^a{}_b) \neq 0$, except on sets of measure zero.) 
While the transformations made in appendices~A.1 and A.2 of reference~\cite{Bobrick:2021} satisfy the determinant condition, they fail the more basic $J^a{}_{[b,c]} =0$ condition; they are simply not coordinate transformations. 

For our purposes then it means that the non-unit-lapse and van~den~Broeck warp drive spacetimes cannot simply be dismissed out of hand. There are other problematic issues with both of these models,  but they are physically different from the Nat\'ario class~\cite{Natario:2001}, (unit lapse, spatially flat 3-geometry), and must be directly addressed using different techniques.

\section{Some defective warp drives}\label{S:defective}

Some of the recently proposed warp drive spacetimes are defective for other reasons.

For instance, in the first explicit example in reference~\cite{Fell:2021} the velocity potential $\Phi$ is certainly $C^0$ but is only piecewise differentiable, $C^{1-}$. That is, the flow $\vec v = \grad \Phi$ is discontinuous, which implies that the metric is discontinuous. This then leads to  delta functions in the Christoffel symbols, squares of delta functions in the Riemann tensor, and cubes of delta functions in the Bianchi identities. This is mathematically and physically not viable. 

In their second  explicit example, equation (9) of versions 1 and 2; equation (11) of versions 3 and 4 of reference~\cite{Fell:2021}, the velocity potential $\Phi$ is $C^\infty$, except possibly at the centre of the spacetime, but suffers other problems. 
If one takes $r=\sqrt{x^2+y^2+z^2}$, 
then even at its most general their velocity potential is of the qualitative form
\begin{equation}
\Phi(x,y,z) =  F(r,\sigma(x,y,z)) + v_* \, z.
\end{equation}
Then
\begin{equation}
\vec v = \grad \Phi(x,y,z) =  \partial_r F(r,\sigma(x,y,z))  \, \hat r 
+ \partial_\sigma  F(r,\sigma(x,y,z)) \, \grad \sigma + v_* \, \hat z.
\end{equation}
But then, in their specific example, they do the equivalent of settting $\sigma(x,y,z)\to1$, and $v_*\to 0$, thereby enforcing spherical symmetry. That is $ \Phi(x,y,z) \to  F(r,1)$,  which is a function of $r$ only. 
But then 
\begin{equation}
\vec v = \grad \Phi 
\to {\partial F(r,1)\over\partial r} \;\;\hat r.
\end{equation}
So their flow vector is always pointing radially outwards/inwards. 
This is simply not viable for describing a warp drive spacetime.

Oddly enough,  consider $\Phi(x,y,z) = 2 \sqrt{2mr} = 2  \sqrt{2m}\;  \sqrt[4]{x^2+y^2+z^2}$, so that
\begin{equation}
\vec v = \grad \Phi(x,y,z) =  {\sqrt{2m} \; (x,y,z) \over (x^2+y^2+z^2)^{3/4}} 
= \sqrt{2m \over \sqrt{x^2+y^2+z^2}} \;\; \hat r.
\end{equation}
This is the Schwarzschild spacetime (in Painlev\'e--Gullstrand form, converted to Cartesian coordinates). This is not a warp drive spacetime. 

Similarly, consider $\Phi(x,y,z) =   {(x^2+y^2+z^2)\over 2\ell}$ so that
\begin{equation}
\vec v = \grad\Phi =  {(x,y,z)\over \ell} = { \sqrt{x^2+y^2+z^2}\over \ell}\; \hat r.
\end{equation}
This is the de Sitter spacetime (in Painlev\'e--Gullstrand form, converted to Cartesian coordinates, with $\Lambda = 3/\ell^2$). This is not a warp drive spacetime. \enlargethispage{20pt}

But the central issue is this: 
Even if you pick a flow vector of the form
\begin{equation}
\vec v = f(r) \; \hat r + v_* \; \hat z,
\end{equation}
which at least has a hope of representing a moving warp bubble, 
then calculating and checking the non-negativity of the Eulerian energy density $\rho$ by itself tells you next to nothing regarding satisfaction of the WEC. 
To test satisfaction of the WEC one would need to calculate all of the principal pressures $p_i$ and then check whether all of the combinations $\rho+p_i$ were non-negative.

\enlargethispage{25pt}
In reference~\cite{Lentz:2020} the claim is made that the author can self-consistently find a warp drive configuration that is sourced by a perfect fluid plasma (satisfying the WEC and so also the NEC) plus electromagnetic Maxwell stress-energy (satisfying all of NEC, WEC, SEC, and DEC).  
So if this claim were to be true, if these models truly were solutions of the Einstein equations with the specified source, one would have a warp drive spacetime satisfying the NEC, which we have just shown to be impossible. What has gone wrong? The point here is that the author of~\cite{Lentz:2020} has not actually solved the Einstein equations, he has only solved part of the Einstein equations --- for the density, flux, and trace of the stress.  This is not enough to obtain a valid solution of the Einstein equations --- the author would also need to consider the remaining trace-free part $T_{ij} -{1\over3} (T_{kl} \delta^{kl}) g_{ij}$ of the stress tensor. 

\section{Final considerations regarding recent warp drive proposals}\label{A:final}

We would like to conclude by explicitly explaining (if not yet mentioned) why each of the recent warp drive proposals have failed in their claims.

Starting with \cite{Lentz:2020}, the author has claimed to have found a warp drive solution which would not violate the WEC. This is actually an incorrect claim since the author has never actually solved the Einstein equations. By imposing a warp drive metric, 
it is clear from equations \eqref{E:Gnn}, \eqref{E:Gni} and \eqref{E:Gij} that \emph{all} of the components of the energy momentum tensor are already determined, without much freedom to play around. This is called a reverse engineering process. Once you define your metric, your energy momentum tensor follows directly from it. The author, however, imposes not only the metric, but also an energy momentum tensor 
--- therefore imposing \emph{both sides} of the Einstein equations without \emph{ever} verifying if they are indeed an equation or not. Hence, they haven't only not found a solution which satisfies the WEC, they haven't found a solution at all.
	
Proceeding with \cite{Bobrick:2021}, their warp drives are divided into two classes: ``General spherically symmetric warp drives'', which we already discussed in appendix \ref{S:spherical}, and ``Axisymmetric warp drives with a general internal region'' in which they haven't found any metric satisfying the WEC. In this way, their only ``warp drive'' which satisfies the WEC is actually not a warp drive metric at all. It is a static spherically symmetric spacetime, therefore being reduced to either Minkowski or Schwarzschild. Furthermore, their claims about Alcubierre's warp drive requiring negative energy due to the ``artificial'' matching rate for the clocks inside and outside the warp drive follows directly from their rediscovery of the existence of time delays in Schwarzschild spacetimes, which they believe to be a warp drive. 

Finally, in \cite{Fell:2021}, their main issue reduces to ``proving'' the WEC by calculating the energy density for a unique class of observers -- the Eulerian observers. As mentioned already, to prove the WEC one needs to prove that the energy density is positive \emph{for all} time-like observers and, proving only for one class of observers simply doesn't prove anything. Furthermore, the specific examples provided by them have a great deal of problems, as already explained in appendix \ref{S:defective}.

\enlargethispage{20pt}
The series of articles by Santos-Pereira, Abreu, and Ribeiro~\cite{Osvaldo-1,Osvaldo-2,Osvaldo-3,Osvaldo-4,Osvaldo-5}, has other issues.
Specifically:
\begin{itemize}
\item 
Their perfect fluid warp drives (dust, non-zero pressure, non-zero pressure with cosmological constant) are just Riemann-flat Minkowski space in disguise. 
\item
Their anisotropic fluid warp drive (``parameterized perfect fluid'' warp drive) is less trivial, but is still seriously diseased.
\item 
Their charged dust warp drive is less trivial, but is still seriously diseased.\\
(Among other things, the calculation of energy density is deeply flawed, confusing $T_{00}$ and $T^{00}$.)
\item
Their ``cosmological'' warp drive is not cosmological.
\end{itemize}
We shall go into some detail in an attempt to make the situation clear and transparent.

The earliest paper~\cite{Osvaldo-1} analyzes the Alcubierre warp drive with 
a stress-energy that is pure comoving Eulerian dust. 
Now this is a very strong constraint, and one quickly deduces that the energy density of the dust is zero. This means that the entire stress-energy tensor is identically zero; so one is looking for \emph{vacuum} solutions
of the Einstein equations for an Alcubierre warp drive.
But the set of vacuum solutions in an Alcubierre warp drive framework is extremely limited --- in fact once one completely satisfies all the Einstein equations the only such solution is Riemann-flat Minkowski space (which, viewed as a warp drive, is trivial). 

The anisotropic fluid (``parameterized perfect fluid'') studied in~\cite{Osvaldo-2} has the defect that once one completely satisfies all the Einstein equations the ``warp drive'' is infinitely wide in the transverse directions --- so it is not a warp ``bubble'', the warp field  stretches all the way across the universe in the $y$ and $z$ directions; one is trying to accelerate an entire ``slab'' of spacetime in the $x$ direction.
Furthermore, the Eulerian energy density, flux, and pressure in the direction of motion are zero, 
the pressures transverse  to the direction of motion are equal and possibly nonzero --- this would then certainly violate the DEC.

For the charged dust solution of~\cite{Osvaldo-3} there is a technical error in identifying the energy density. The physically interesting quantity is $\rho = T^{00} = T^{ab}\; (-1;0,0,0)_a \; (-1;0,0,0)_b$, where the 4-covector $(-1;0,0,0)_a$ is always timelike and future pointing (it is in fact the 4-velocity of an Eulerian observer). In contrast the quantity  $T_{00} = T_{ab}\; (1;0,0,0)^a \; (1;0,0,0)^b$ is not physically interesting since the 4-vector $(1;0,0,0)^a$ is timelike only for a  limited range of warp velocities; in fact, this 4-vector becomes spacelike exactly when the warp bubble motion becomes superluminal. 

For the ``cosmological'' solution of~\cite{Osvaldo-4} the attempt at including a cosmological constant is at best a redefinition of terms in the stress-energy tensor. For instance, the definition of Einstein tensor adopted in~\cite{Osvaldo-4} is not the usual one, which is unhelpful.
Furthermore, given that the entire calculation is taking place within the framework of the generic Alcubierre metric, we know the spacetime is asymptotically flat; it does not match (anti)de~Sitter at large distances, 
so it is not a ``cosmological'' warp bubble.

In all of these situations~\cite{Osvaldo-1,Osvaldo-2,Osvaldo-3,Osvaldo-4,Osvaldo-5},
one is working within the framework of the generic Alcubierre metric,
so we know from first principles that energy condition violations must be present.

\clearpage
\section*{Acknowledgements}
JS  acknowledges indirect financial support via the Marsden fund, 
administered by the Royal Society of New Zealand.\\
SS acknowledges financial support via OP RDE project No. CZ.02.2.69/0.0/0.0/18\_053/\-0016976 (International mobility of research), and the technical and administrative staff at the Charles University.\\
MV was directly supported by the Marsden Fund, via a grant administered by the Royal Society of New Zealand. \\
The authors would like to thank the anonymous referee for making several very practical suggestions, helping us to improve the presentation and logic flow in this article.

\bigskip
\hrule

\end{document}